\documentclass[lettersize,journal]{IEEEtran}
\usepackage{amsmath,amsfonts}
\usepackage{algorithmic}
\usepackage{enumitem}
\usepackage[linesnumbered,ruled,vlined]{algorithm2e}
\usepackage{setspace}
\usepackage{array}
\usepackage{textcomp}
\usepackage{stfloats}
\usepackage{multirow,diagbox,tabularx,blindtext}
\usepackage{booktabs}
\usepackage{hyperref}
\usepackage{verbatim}
\usepackage{graphicx}
\usepackage{amssymb}
\usepackage{textcomp}
\usepackage{cite}
\usepackage{color}
\usepackage{pifont}
\usepackage{makecell}
\usepackage{arydshln}
\usepackage{mathrsfs}   
\usepackage[T1]{fontenc}
\usepackage[scaled=0.81]{beramono}

\newcommand{\PP}[1]{
\vspace{2px}
\noindent{\bf \IfEndWith{#1}{.}{#1}{#1.}}
}

\usepackage{subfigure}
\usepackage[normal]{caption}
\usepackage{enumitem}
\usepackage{colortbl}
\definecolor{bgreen}{RGB}{0,170,0}
\definecolor{bred}{RGB}{220,0,0}
\definecolor{mydarkblue}{RGB}{0,0,150}
\definecolor{Gray}{gray}{0.93}
\hypersetup{
    colorlinks=true,
    linkcolor=bred,
    citecolor=mydarkblue,
    filecolor=bred,
    urlcolor=mydarkblue
}

\begin{document}

\begin{sloppypar}

\title{\huge KGBERT4Eth: A Feature-Complete Transformer Powered by Knowledge Graph for Multi-Task Ethereum Fraud Detection} 

\author{Yifan Jia, Ye Tian, Liguo Zhang, Yanbin Wang*, Jianguo Sun*, Liangliang Song

\thanks{\emph{(* Corresponding author: Yanbin Wang, Jianguo Sun.)} }       

\thanks{Yifan Jia is with the Yantai Research Institute, Harbin Engineering University, Yantai 264000, China (e-mail:jiayf@hrbeu.edu.cn). 

Ye Tian, Liangliang Song and Jianguo Sun are with the Hangzhou Research Institute, Xidian University, Hangzhou 311231, China (e-mail:tianye@xidian.edu.cn;songliangl@xidian.edu.cn;jgsun@xidian.edu.cn).

Liguo Zhang is with the College of Computer Science and Technology, Harbin Engineering University, Harbin 150001, china (e-mail:zhangliguo@hrbeu.edu.cn)

Yanbin Wang is with the College of Computer Science and Technology, Zhejiang University, Hangzhou 310058, China (e-mail:wangyanbin15@mails.ucas.ac.cn). }
}

\markboth{IEEE Transactions on Knowledge and Data Engineering} 
{Jia \MakeLowercase{\textit{et al.}}: KGBERT4Eth: A Feature-Complete Transformer Powered by Knowledge Graph for Multi-Task Ethereum Fraud Detection} 


\maketitle

\begin{abstract}
Ethereum's rapid ecosystem expansion and transaction anonymity have triggered a surge in malicious activity. Detection mechanisms currently bifurcate into three technical strands: expert-defined features, graph embeddings, and sequential transaction patterns - collectively spanning the complete feature sets of Ethereum's native data layer. Yet the absence of cross-paradigm integration mechanisms forces practitioners to choose between sacrificing sequential context awareness, structured fund-flow patterns, or human-curated feature insights in their solutions. To bridge this gap, we propose KGBERT4Eth, a feature-complete pre-training encoder that synergistically combines two key components: (1) a Transaction Semantic Extractor, where we train a enhanced Transaction Language Model (TLM) to learn contextual semantic representations from conceptualized transaction records, and (2) a Transaction Knowledge Graph (TKG) that incorporates expert-curated domain knowledge into graph node embeddings to capture fund flow patterns and human-curated feature insights. We jointly optimize pre-training objectives for both components to fuse these complementary features, generating feature-complete embeddings. Notably, to emphasize rare anomalous transactions, we design a biased masking prediction task for TLM to focus on statistical outliers, while the Transaction TKG employs link prediction to learn latent transaction relationships and aggregate knowledge. Furthermore, we propose a mask-invariant attention coordination module to ensure stable dynamic information exchange between TLM and TKG during pre-training. KGBERT4Eth significantly outperforms state-of-the-art baselines in both phishing account detection and de-anonymization tasks, achieving absolute F1-score improvements of 8–16\% on three phishing detection benchmarks and 6–26\% on four de-anonymization datasets. Our source code can be found at https://anonymous.4open.science/r/KGBERT4ETH-0BB5.
\end{abstract}


\section{Introduction}

Blockchain has gone through the blockchain 1.0 form represented by the digital cryptocurrency Bitcoin, and entered the blockchain 2.0 form dominated by smart contracts. Ethereum, as one of the most popular and scalable blockchains, has a market value of more than \$450 billion to date. However, the increasing popularity and value of Ethereum has attracted malicious actors who intend to use the platform for economic gain \cite{ghosh2025catalog,li2025blockchain}. They use the high anonymity and immutability of transactions on the blockchain to deceive users or systems for profit through transaction manipulation and fake dApps \cite{zhou2023dapphunter,chen2024dissecting,wu2023toward_intro}.

Ethereum fraud detection primarily involves learning detection models from historical transaction data. Existing learning approaches can be categorized into three distinct paradigms based on feature processing methodologies: expert-defined feature models, graph embedding learning models, and sequential transaction modeling. The first paradigm employs manually-designed statistical features from transaction metadata and network structures to identify suspicious behavioral signatures. Graph-based methods construct transactional networks to learn topological representations of money flow patterns, while sequential approaches model contextual relationships in transaction sequences. Although these complementary perspectives form a theoretically complete feature space, their effective integration is nontrivial - demanding sophisticated cross-paradigm coupling mechanisms, particularly between graph and sequence representations, to achieve true endogenous feature fusion during model training rather than simplistic late-stage feature concatenation.

Furthermore, beyond the challenges of feature completeness and fusion methodologies, existing approaches face several granular limitations in their implementations: 

\begin{itemize}

    \item Expert-defined features: Current methods predominantly focus on isolated account-level characteristics (e.g., nodal in/out degrees, transfer frequencies) and limited network metrics (e.g., degree/betweenness centrality), while overlooking temporal activity indicators such as account lifespan and active days. The absence of comprehensive feature engineering encompassing three dimensions - transaction statistics, temporal activity patterns, and network structural metrics - represents information loss.
    
    \item Graph embedding: Graph embedding-based detection methods typically model accounts as static nodes with predefined attributes, potentially introducing representation bias. Furthermore, most existing approaches rely on neighborhood aggregation mechanisms - a learning paradigm that often obscures distinctive features of rare anomalous nodes in Ethereum's large-scale networks where malicious entities constitute a minority population \cite{chen2020measuring_graph_smooth,keriven2022graph_smooth}.
    
    \item Sequence modeling: Recent advances in sequence modeling leverage pre-trained Transformers, including BERT4ETH \cite{bert4eth,zipzap} and Transaction Language Model TLMG4Eth \cite{sun2025ethereum, sheng2025dynamic}. Although TLMG4Eth advances the field through numerical record conceptualization for richer semantic representation, its handling of Ethereum's transaction homogeneity is insufficient. Furthermore, existing sequence models ignore gas-related transactional attributes that could provide crucial behavioral signals.
\end{itemize}

To address these limitations, we propose KGBERT4Eth—a feature-complete pre-training framework that couples semantic, structural, and expert-defined signals. The architecture comprises two synergistic components: (1) a Transaction Semantic Extractor featuring our enhanced Transaction Language Model (TLM) for capturing contextual dependencies and transaction semantics, and (2) a Transaction Knowledge Graph (TKG) that embeds multifaceted domain knowledge (including fund flows, inter-entity relationships, and expert-curated behavioral features) into graph-node representations. KGBERT4Eth uses dual mechanisms for effective feature integration: First, it uses joint optimization of TML and TKG pretraining objectives, where TML leverages our designed biased masked prediction task to reduce focus on homogeneous transaction records, and TKG employs link prediction to uncover latent transaction associations and aggregate structured knowledge. Second, we propose a mask-invariant attention synergy module to preserve the integrity of the language model's masked semantic objectives during information exchange in joint pretraining, enabling conflict-free cross-modal embedding alignment.

Our key contributions are summarized as follows:
\begin{itemize}[leftmargin=*]

\item We propose KGBERT4Eth, a feature-complete pre-trained fraud detection framework that significantly outperforms 15 baselines across two core Ethereum fraud detection tasks, achieving absolute F1-score improvements of 8–16\% for phishing account detection and 6–26\% for de-anonymization.

\item We present an enhanced Ethereum transaction language model that  advances transaction modeling in two aspect: (i) an expanded feature space encompassing gas metadata and other previously omitted transaction attributes, and (ii) a specialized biased masking mechanism that counteracts the representation bias caused by homogeneous transaction patterns, resulting in more robust semantic learning.

\item We construct an Ethereum transaction knowledge graph that formally defines multiple account types, encodes the fund flow network, and embeds expert-curated behavioral features, thereby offering significantly more comprehensive knowledge encoding.

\item We propose a joint optimization approach addresses cross-paradigm integration through: (i) simultaneous training of TLM and TKG under unified objectives, and (ii) a mask-invariant attention module that preserves the integrity of masking objectives during inter-model information exchange - ensuring stable feature alignment, effectively resolving the representation conflicts between TLM and TKG.
\end{itemize}

\section{Backgroud and Related work}
\subsection{Ethereum Architecture and Account}
Ethereum is a decentralized blockchain platform that enables the execution of smart contracts and the development of decentralized applications (dApps) \cite{chen2018detecting_bkg}. Its account model is structured around two distinct types: externally owned accounts (EOAs) and contract accounts.

EOAs are user-controlled entities, secured by cryptographic private keys. They can initiate transactions directly, including the transfer of Ether and the invocation of contract functions \cite{tabatabaei2023understanding}. As they are controlled by identifiable users or organizations, EOAs are often implicated in security-related incidents, such as phishing attacks and unauthorized asset transfers \cite{luo2024ai_bkg,yang2024tifs_bkg}. Consequently, they serve as a primary analytical focus in fraud detection systems.

Contract accounts, on the other hand, are associated with smart contracts deployed on-chain. These accounts do not possess private keys and cannot autonomously initiate transactions. Instead, they operate reactively, executing predefined logic only when invoked by external transactions originating from EOAs. Although passive in isolation, contract accounts are frequently utilized in malicious schemes. Adversaries may deploy deceptive smart contracts designed to exploit user interactions, thereby automating the misappropriation of assets or concealing transaction trails. These mechanisms are commonly observed in fraudulent dApps and obfuscation-based fund laundering \cite{xie2024defort,liang2025towards}. The interaction between EOAs and contract accounts forms a core structural component of Ethereum’s transaction ecosystem. 

\subsection{Related Work}

Existing research \cite{he2023txphishscope_ccs,guan2024characterizing_ccs,wen2023novel_hybrid_feature_graph_related_work,hu2021transaction_based_related,jin2024Contrastive_Self-supervision_tifs,yang20242dynethnet_tifs,liang2025tosem,ghosh2025treat} efforts have extensively explored the detection of fraudulent activities on Ethereum by leveraging both traditional and graph-based machine learning approaches. Early studies primarily focused on account-level statistical patterns extracted from transaction records. For instance, Chen et al. introduced a dual-sampling ensemble method \cite{chen2020ijcai_lgbm} based on LightGBM, incorporating cascading features derived from transaction histories to improve phishing detection accuracy. Xu et al. proposed a framework \cite{xu2024ewdps_related_work} for modeling the dynamic evolution of phishing scams, aiming to facilitate early-to-mid-stage fraud warning based on behavioral sequences. Traditional approaches demonstrated effectiveness in capturing localized transactional anomalies but often struggled to generalize in large-scale, complex interaction networks.

Given that Ethereum transactions naturally form graph-structured data, a considerable body of research has adopted graph-based methods to model account interactions \cite{zhou2022behavior_tifs_deanoy,xia2022graph_related_work,liu2024gang,chen2020phishing_TOIT_graph_related_work}. Trans2Vec~\cite{trans2vec}, built on DeepWalk~\cite{deepwalk}, guides biased random walks using transaction amounts and timestamps to learn structure-aware embeddings of accounts. Lin et al.~\cite{lin2023phish2vec} extended this by considering both structural and transactional homogeneity in guiding the embedding process, thereby improving similarity modeling between accounts. Further advancements integrate graph neural networks (GNNs) with machine learning techniques to enhance fraud detection performance. For example, TGC~\cite{GNN_r3} employs subgraph contrastive learning with statistical features to identify phishing accounts. TGAT-based methods~\cite{wang2023tgat} exploit attention mechanisms to capture temporal transaction dynamics, while SIEGE~\cite{li2023siege} introduces self-supervised incremental learning tasks to improve model adaptivity over time. More recently, TokenScout~\cite{wu2024tokenscout} models temporal graphs for fraud detection in token trading, and TSGN~\cite{tsgn} integrates handcrafted features with DiffPool-based subgraph representations. GrabPhisher~\cite{zhang2024grabphisher} captures evolving temporal and topological features by constructing continuous-time diffusion graphs of account behavior.

Despite their effectiveness in modeling topological structures and local interactions, GNN-based methods exhibit inherent limitations. They generally treat accounts as static nodes with fixed features, which restricts the ability to model evolving behavioral patterns. Moreover, in large-scale networks where malicious accounts are rare, neighborhood aggregation may dilute discriminative signals, leading to over-smoothing and reduced detection sensitivity. To address these challenges, recent studies have explored Transformer-based architectures for encoding Ethereum transaction records. Models such as BERT4ETH~\cite{bert4eth} and ZipZap~\cite{zipzap} adopt pre-trained language models to capture sequential dependencies in numerical transaction features. However, their use of context-agnostic input limits the modeling of semantic relationships and complex behavioral patterns. To enhance contextual understanding, Sun et al.~\cite{sun2025ethereum} and Sheng et al.~\cite{11022733} proposed treating transaction records as textual sequences for semantic modeling. While this language-centric paradigm offers a promising alternative, it remains constrained by the high structural homogeneity of Ethereum transactions and the absence of structured domain knowledge, both of which are essential for deep semantic understanding in blockchain settings.


\section{OUR INSIGHTS}
Despite recent advances in Ethereum fraud detection using language models and graph learning, several technical challenges remain unresolved. (1) Language models pre-trained on general text corpora often struggle to capture the latent semantics of highly homogeneous yet subtly anomalous transaction sequences. (2) Transformer-based encoders focus primarily on sequential dependencies, neglecting the structured relational context that is critical for characterizing inter-account behaviors. (3) Existing models that attempt to incorporate structural knowledge often rely on explicit graph construction at inference time , which limits scalability and fails to bridge semantics with structural priors during learning. These limitations motivate the following three insights, which form the foundation of our KGBERT4Eth framework.

\begin{itemize}
    \item \textbf{Key Insight 1: Homogeneous transaction texts require task-aware pre-training objectives to enhance semantic discriminability.} 
    Ethereum transaction records exhibit structured formats and frequent field repetition, which diminish the effectiveness of conventional masked language modeling (MLM) objectives that randomly mask tokens without considering their semantic contribution. To address this, we propose a biased masking strategy, which prioritizes the masking of tokens with high informativeness, as measured by their statistical rarity and relevance within each account’s transaction corpus. This encourages the model to focus on semantically salient components, enhancing its capacity to detect irregular transaction patterns while maintaining contextual coherence for token prediction.

    \item \textbf{Key Insight 2: Transactional structure and domain-specific account features are essential for capturing relational semantics beyond sequential modeling.}
    Fraudulent behaviors often emerge not from isolated transaction content but from relational interactions among accounts. To capture such patterns, we construct a Transaction Knowledge Graph (TKG) that encodes inter-account relations using heterogeneous edges and node-level expert features. These features include transaction frequencies, statistic features, and multiple graph-theoretic centrality scores. Unlike text-only models, this structure-aware representation enables the model to learn aggregated behavioral profiles, offering essential inductive biases for downstream detection tasks.

    \item \textbf{Key Insight 3: Integrating structured knowledge during pre-training through cross-modal coordination, and avoid semantic conflicts.}
    While fusing semantic and structural information is beneficial, premature integration may undermine the language model’s core objective of masked token prediction (MP). Injecting knowledge representations too early may corrupt or overwrite masked positions, reducing the effectiveness of semantic learning. A common alternative is to pre-train the semantic and structural modules separately, allowing each modality to specialize before their representations are combined. However, such independent training and post hoc fusion fail to enable mutual enhancement during pre-training and often require reconstructing structural representations during inference. To resolve this conflict, we design an attention synergy mechanism that preserves the integrity of masked tokens during learning, enabling structured knowledge to enhance semantic embeddings without overwriting or distorting the core language objectives. By learning to integrate graph-derived knowledge with token-level semantics during pre-training, the model retains the capacity to implicitly utilize relational knowledge during inference, even in the absence of explicit graph representations.
\end{itemize}

In light of the above, our work builds upon prior advances by introducing a knowledge-grounded dual-encoder framework that unifies semantic learning from transaction texts and structured knowledge modeling via a transaction knowledge graph. Unlike previous methods, we integrate contextualized language modeling with expert-curated knowledge representations through a synergy mechanism that maintains information consistency and minimizes semantic conflicts. This enables the model to effectively encode both fine-grained transactional semantics and high-level relational dependencies, which are essential for identifying subtle anomalies and behaviorally coordinated fraud on Ethereum.

\begin{table}[t]
\small
\centering
\caption{The Ethereum Transaction (tx) Fields and Simplified Descriptions.}
\setlength{\tabcolsep}{1mm}
\renewcommand{\arraystretch}{1.11}
\begin{tabular}{ll}
\noalign{\hrule height 1.2pt}
\textbf{Field} & \textbf{Description} \\ \hline
blockNumber & Block height of the transaction. \\
timeStamp & Time when the tx was confirmed. \\
hash & Unique identifier of the transaction. \\
nonce & Sender’s transaction count. \\
blockHash & Hash of the containing block. \\
transactionIndex & Index of the tx within the block. \\
from & Address of the transaction sender. \\
to & Address of the recipient. \\
value & Amount of Ether transferred, in Wei. \\
gas & Maximum gas allowed for execution. \\
gasPrice & Fee per unit of gas, specified in Wei. \\
isError & Indicates transaction failure or success. \\
txreceipt\_status & Status code of the transaction receipt. \\
contractAddress & Contract address created, if applicable. \\
cumulativeGasused & Gas used in the block up to this tx. \\
gasUsed & Actual gas consumed by this tx. \\
confirmations & Num of blocks confirming the tx. \\
methodId & Encoded identifier of the called method. \\
functionName & Name of the contract invoked invoked. \\
\noalign{\hrule height 1.2pt}
\end{tabular}
\label{tab:eth_tx_fields}
\end{table}

\section{Preliminary}
\subsection{Ethereum Account Transaction Records}
\label{sec2.1}
An Ethereum account transaction record consists of transactions an account participates in, denoted as \( \hat{T}_i = \{t_1, t_2, \dots, t_n\} \), where \( n \) is the total number of transactions. Each transaction includes multiple fields, \( \{e_1, e_2, \dots, e_{|E|}\} \in E \), such as \texttt{blockNumber}, \texttt{timeStamp}, and \texttt{transactionIndex}, which are publicly accessible on the blockchain. We provide an overview of these native transaction fields in Table~\ref{tab:eth_tx_fields}, which summarizes the core on-chain attributes available for each Ethereum transaction. To capture meaningful semantic information, we exclude fields that lack inherent features. For example, the \texttt{from} and \texttt{to} fields store Ethereum addresses, which are 40-character hexadecimal strings serving only as identifiers, making them difficult for language models to tokenize effectively. Instead, we retain five essential fields: \texttt{value}, \texttt{timeStamp}, \texttt{IO}, \texttt{gas}, and \texttt{gasPrice}, denoted as \( \hat{e}_1, \hat{e}_2, \dots, \hat{e}_{|\hat{E}|} \in \hat{E} \). For the interaction patterns implicitly revealed by Ethereum addresses, we adopt a more suitable graph-based structure in subsequent sections to explicitly capture and model such relationships. 

\subsection{Transaction Knowledge Graph (TKG)}

We construct the Transaction Knowledge Graph (TKG) from Ethereum transaction records to capture the extensive relationships between accounts, addressing the limitations of sequence models that fail to incorporate detailed transaction addresses. The knowledge graph is represented as a multi-relational graph $G = (V, E)$, where $V$ is the set of entity nodes and $E \subseteq V \times R \times V$ is the set of edges. These edges connect nodes in $V$ via triples, with $R$ being the set of relationship types. Each triple is represented as $(h, r, t)$, where $h, t \in V$ and $r \in R$, denoting the transactional relationships within the knowledge graph. In our constructed TKG, each entity represents an Ethereum address, including both external and contract accounts (EOA and CA). The relationship types, $r$, are divided into two categories: external transactions and contract transactions, which indicate whether the transaction between two entities is a regular transfer or a transfer through a smart contract. 

\subsection{Expert Knowledge}

\begin{table*}[ht]
\small
\centering
\renewcommand{\arraystretch}{1.1}
\caption{Expert-engineered features used to initialize nodes in the Transaction Knowledge Graph.}
\setlength{\tabcolsep}{7mm}  
\begin{tabular}{ll}
\noalign{\hrule height 1.5pt}
\textbf{Feature} & \textbf{Description} \\ \hline
Node outdegree & Number of outgoing transactions from the account. \\ 
Node indegree & Number of incoming transactions to the account. \\ 
Direction ratio & Ratio of incoming to outgoing transactions. \\ 
Max outgoing amount & Maximum transaction amount sent by the account. \\ 
Min outgoing amount & Minimum transaction amount sent by the account. \\ 
Max incoming amount & Maximum transaction amount received by the account. \\ 
Min incoming amount & Minimum transaction amount received by the account. \\ 
Average outgoing amount & Average transaction amount sent by the account. \\ 
Average incoming amount & Average transaction amount received by the account. \\ 
Account balance & The sum of all income minus the sum of outgoing. \\ 
Account lifetime & Duration of time the account has been active. \\ 
Active days & Number of distinct days with at least one transaction. \\ \hdashline
Long-term transfer frequency & Number of transactions over a long-term window. \\
Short-term transfer frequency & Number of transactions over a short-term window. \\ 
Long-term incoming transfer frequency & Number of incoming transactions over a long-term window. \\ 
Short-term incoming transfer frequency & Number of incoming transactions over a short-term window. \\ 
Long-term outgoing transfer frequency & Number of outgoing transactions over a long-term window. \\ 
Short-term outgoing transfer frequency & Number of outgoing transactions over a short-term window. \\ \hdashline
Katz centrality & Influence of an account considering all possible paths to it. \\ 
Betweenness centrality & Frequency of an account appearing in shortest paths between others. \\ 
Degree centrality & Number of direct connections an account has in the graph. \\ 
Closeness centrality & Average shortest path length from an account to all other accounts. \\ 
Clustering coefficient & Degree to which an account is embedded in a local transaction network. \\ 
Eigenvector centrality & Importance of an account based on the influence of its neighbors. \\ 
Indegree centrality & Number of incoming connections to an account. \\ 
Outdegree centrality & Number of outgoing connections from an account. \\ 
\noalign{\hrule height 1.5pt}
\end{tabular}
\renewcommand{\arraystretch}{1.2}
\label{tab:node_features}
\end{table*}


\begin{figure}[ht]
    \centering
    \begin{minipage}[b]{0.241\textwidth}
        \centering
        \includegraphics[width=1\textwidth]
        {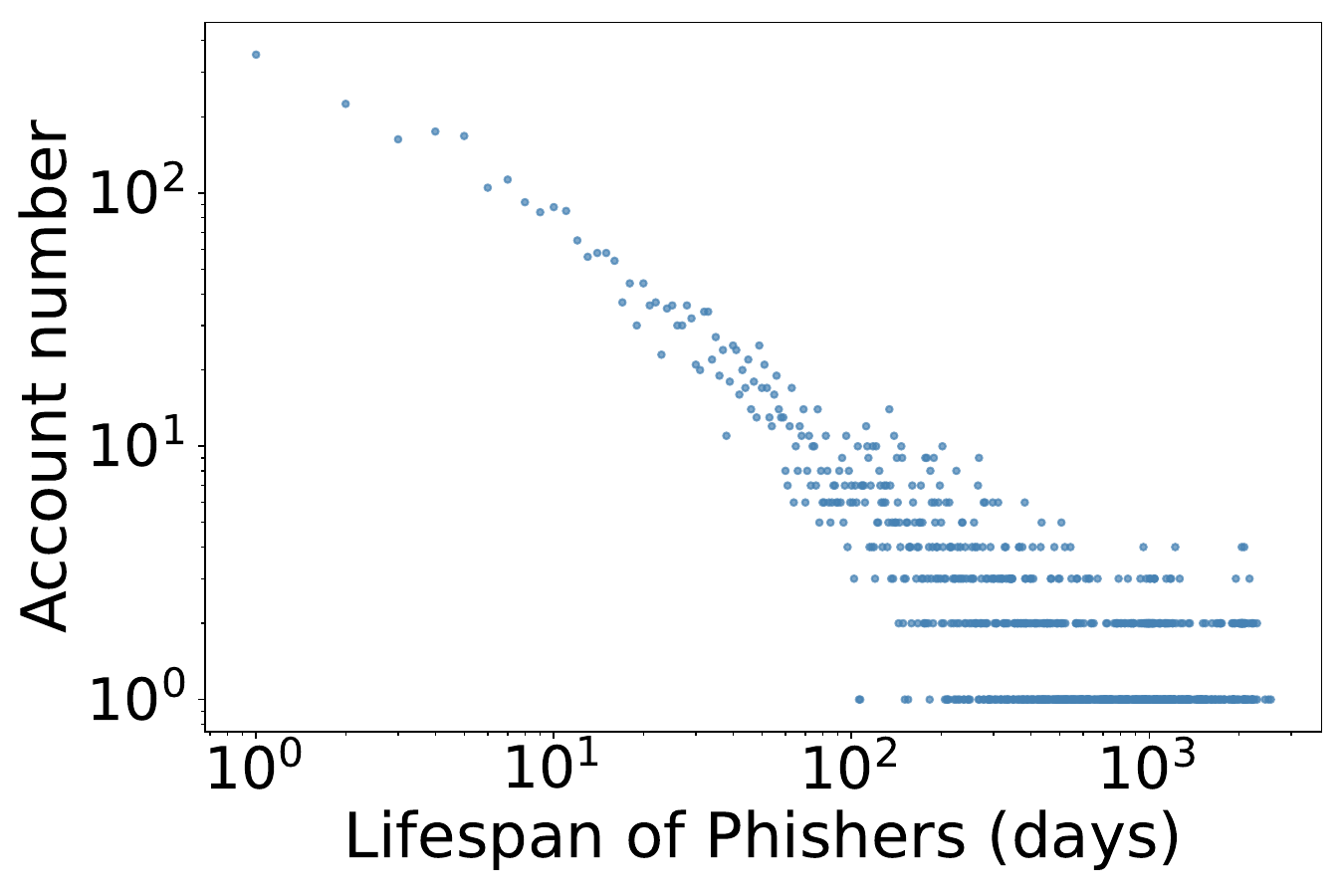}
    \end{minipage}
    \begin{minipage}[b]{0.241\textwidth}
        \centering
        \includegraphics[width=1\textwidth]{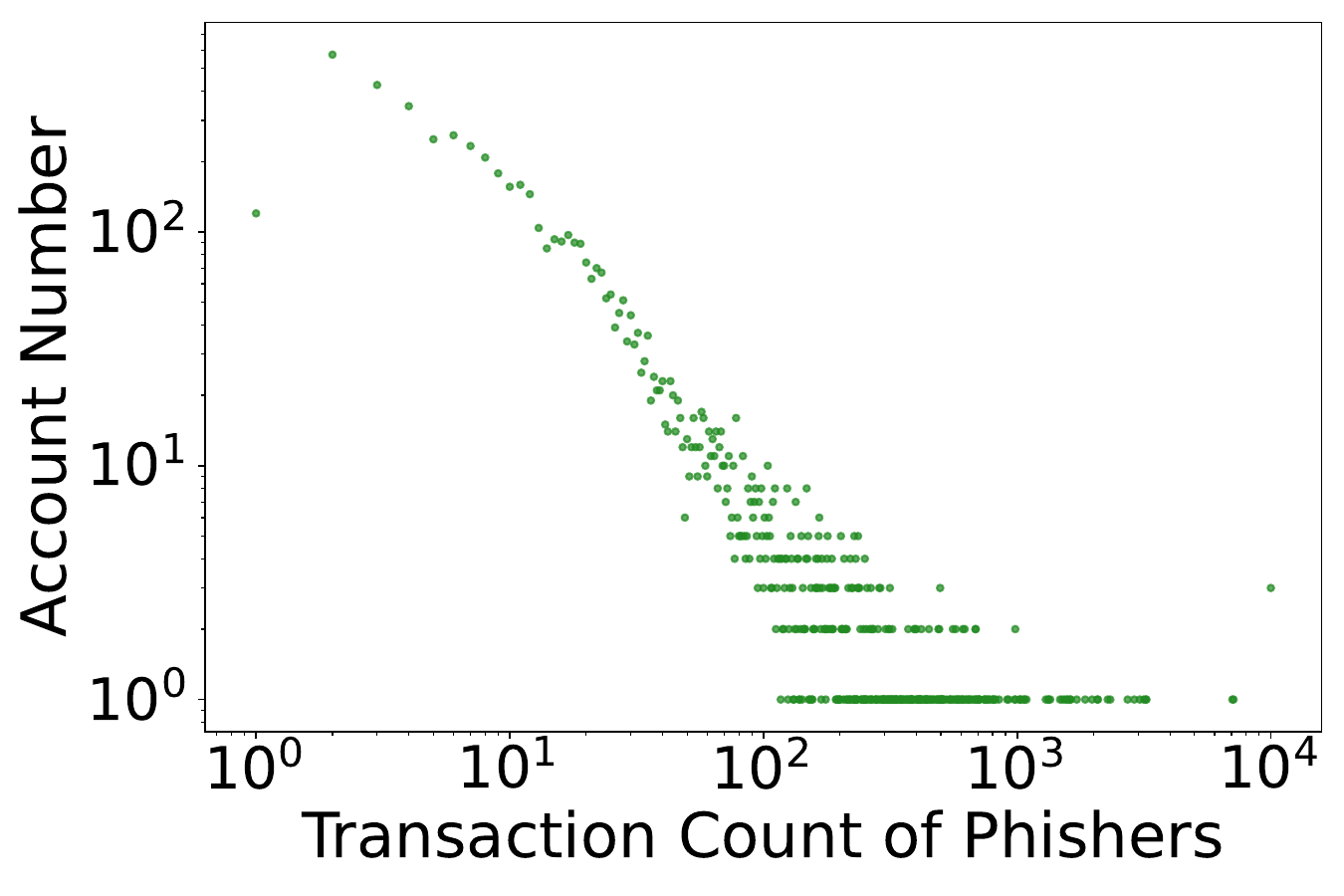}
    \end{minipage}
    \caption{Empirical distribution of phishing accounts in terms of lifespan (\emph{left}) and transaction count (\emph{right}). Both statistics exhibit power-law distribution, suggesting that most phishing accounts are short-lived and conduct few transactions, while a small fraction persist longer or act more aggressively.}
    \label{fig:powerlaw distribution}
\end{figure}

We perform quantitative analysis on each account's transaction records and design expert-engineered features to initialize the entity embeddings in the Transaction Knowledge Graph. These features are intended to capture diverse behavioral signals that distinguish normal accounts from those involved in phishing and other illicit activities. As shown in Table~\ref{tab:node_features}, the features encompass three main categories: statistical descriptors of financial behavior, temporal activity patterns, and structural network metrics. Statistical features such as transaction volume extrema, directionality ratios, and net balance provide coarse-grained indicators of account-level economic activity. These are particularly useful for identifying anomalies such as disproportionately large transfers or abrupt inflow-outflow imbalances. Temporal features, including short-term and long-term transfer frequencies, account lifespan, and active days, reflect the dynamic behavior of accounts over time and are effective in capturing bursty transaction patterns. Structural features derived from the transaction graph encode connectivity and influence, leveraging centrality and clustering metrics to capture roles that accounts play within the broader network topology.

Empirical distributions of two representative temporal features are depicted in Fig.~\ref{fig:powerlaw distribution}. As illustrated in figure, both the lifespan and the transaction-count curves exhibit pronounced power-law distribution: a small fraction of phishing accounts remain active for more than one year or execute thousands of transfers, whereas the vast majority operate briefly and sparsely.This indicates that most phishing accounts exhibit extremely short activity periods and low transaction counts, while a small number remain active for prolonged durations or conduct large numbers of transfers. These findings support the use of heavy-tailed statistical indicators in feature design and suggest that incorporating such priors can improve the model's ability to capture atypical behaviors. By embedding these carefully designed features into each node of the knowledge graph, we enable the model to encode rich semantic and topological context in a unified representation space, thereby facilitating more effective downstream fraud detection.


\section{Method}

\begin{figure*}[ht]
\centering
\includegraphics[width=1\textwidth]{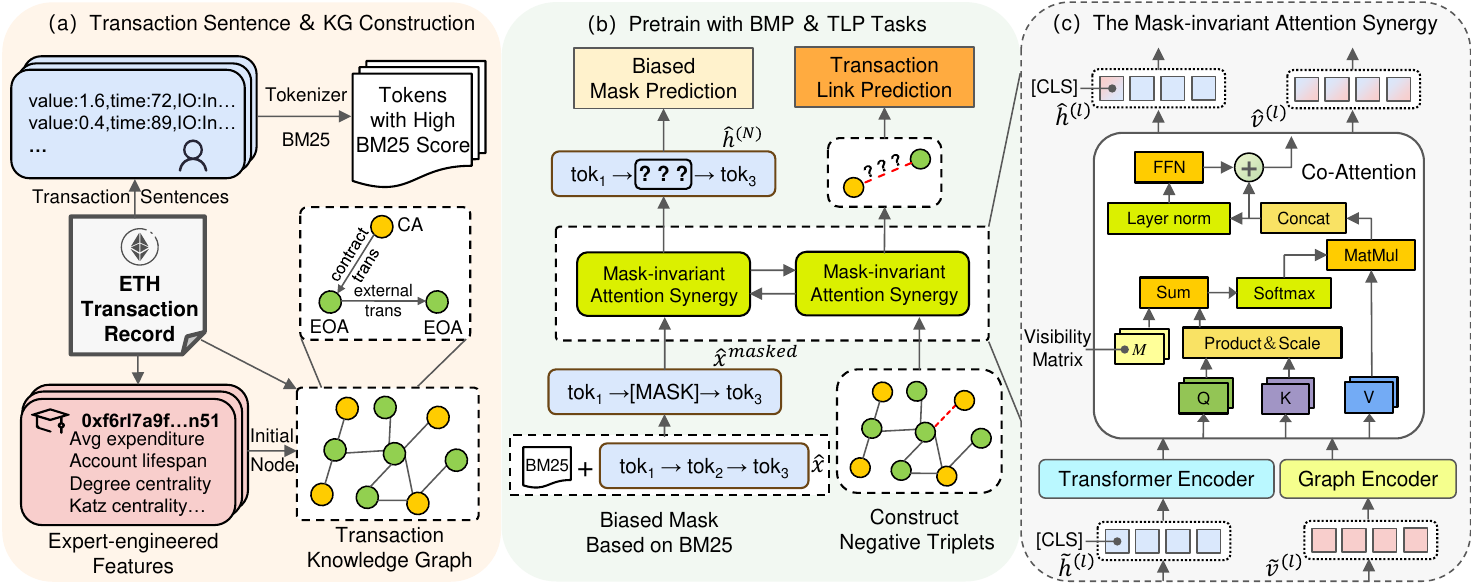}
\caption{An overview illustration of model. (a) Construct transaction sentences and TKG from Ethereum transaction records, using expert features to initialize the TKG. (b) The model is pre-trained by synergizing transaction sentences with corresponding TKG through the proposed mask-invariant attention synergy method. Two pre-training tasks are unified to learn contextualized representations. (c) The mask-invariant attention synergy based on co-attention mechanism.}
\label{fig:framework}
\end{figure*}

In this section, we introduce the architecture of our proposed model (Figure. \ref{fig:framework}) and the detailed steps involved in processing the data.

\subsection{Data Processing}
\label{data processing}

\paragraph{\textbf{Textualization of Transaction Records}}

To process the numerical transaction records of an account $i$, denoted as $\hat{T}_i = \{t_1, t_2, \dots, t_n\}$, into a linguistically interpretable representation, we textualize $\hat{T}_i$ along with their field names $\hat{E}$. The textual representation of the account's transaction record is denoted as $\hat{X}_i = \{x_1, x_2, \dots, x_n\}$. For the $k$-th transaction record of account $i$, where $k \in [1, n]$, its textualized form can be expressed as follows example:
\begin{equation}
x_k=\texttt{str}(
    \small\text{value:0.3, }
    \small\text{timeStamp:14, }
    \small\text{IO:0, }
    \small\text{gas:21, }
    \small\text{gasPrice:3.1})
\label{eq:textualized_record}
\end{equation}

Here, the function \texttt{str()} converts each transaction field and its corresponding value into a readable string format, combining field names from $\hat{E}$ and their records values.

After converting $\hat{T}_i$ into $\hat{X}_i$, we apply a BERT tokenizer to obtain a collection of tokenized words $w_i$ for account $i$’s textualized transaction records $\hat{X}_i$. By repeating this process for all accounts, we build a comprehensive  Ethereum transaction corpus \(\mathcal{C}=\{\hat{X}_1,\hat{X}_2,\dots,\hat{X}_{N_a}\}\) and a vocabulary $\mathcal{D} = \{\hat{E}, W\}$, where \(N_a\) is the total number of accounts, $W$ contains the union of tokenized words from all accounts, i.e., $W = w_1 \cup w_2 \cup \dots \cup w_{N_a}$.

\paragraph{\textbf{Biased Mask Strategy}}
General natural language models struggle to process the textualized transaction records presented in Equation~\eqref{eq:textualized_record}, often failing to generate discriminative embeddings for accounts. To address this limitation, we propose a biased masking method based on BM25, which prioritizes the most representative tokens in the transaction text of each account for subsequent pre-training tasks.

The BM25 score, calculated for each token \( tok_j \in w_i \) in the transaction text \( \hat{X}_i \) of an account, serves as the foundation of our masking strategy. Specifically, the BM25 calculation is based on the following equations:
\begin{equation}
\begin{aligned}
\text{IDF}(tok_j) &= \log \frac{|\mathcal{C}|}{|\{ \hat{X}_i \in \mathcal{C} : tok_j \in \hat{X}_i \}| + 1}, \\
\text{BM25}(tok_j, \hat{X}_i) &= \frac{f(tok_j, \hat{X}_i) \cdot (\text{IDF}(tok_j) \cdot (z_1 + 1))}{f(tok_j, \hat{X}_i) + z_1 \cdot \left(1 - b + b \cdot \frac{|\hat{X}_i|}{avgdl}\right)}.
\end{aligned}
\label{eq:bm25}
\end{equation}
Here, \( \text{IDF}(tok_j) \) represents the inverse document frequency of the token \( tok_j \), which reflects the rarity of the token in the corpus \( \mathcal{C} \). Intuitively, if a token is rare across the entire corpus but appears frequently in a specific document, it is considered more discriminative for that document, and its corresponding BM25 score will be higher. The term \( f(tok_j, \hat{X}_i) \) denotes the frequency of \( tok_j \) in the transaction text \( \hat{X}_i \), while \( |\hat{X}_i| \) and \( {avgdl} \) represent the length of \( \hat{X}_i \) and the average length of transaction texts in the corpus, respectively. The \( z_1 \) and \( b \) control the sensitivity of BM25 to token frequency and text length.

Based on these calculations, we determine the importance of each token within \( \hat{X}_i \). To incorporate this insight into pre-training, we implement a biased masking strategy where a token \( tok_j \) is selected for masking if its normalized BM25 score exceeds a predefined threshold \( \tau \). The masking condition is defined as:
\begin{equation}
\text{Mask}(tok_j) = 
\begin{cases}
1 & \text{if } \displaystyle\frac{\text{BM25}(tok_j, \hat{X}_i)}{\sum_{tok_j' \in \hat{X}_i} \text{BM25}(tok_j', \hat{X}_i)} > \tau, \\
0 & \text{otherwise}.
\end{cases}
\label{eq:mask_threshold}
\end{equation}
Here, \( \text{Mask}(tok_j) = 1 \) indicates that the token \( w \) is selected for masking, and \( \tau \) is a hyperparameter that determines the threshold for masking. This approach ensures that only tokens with significant importance in the transaction text are masked, while less representative tokens are retained to provide contextual information for recovering these masked tokens, guiding the model to focus on reconstructing the most informative and distinguishing parts of the transaction text. 

\subsection{Data Encoding}

\paragraph{\textbf{Transaction Sequence Encoder}}

The pre-processed textual transaction records of an account, after undergoing biased masking, are encoded using an $N$-layer Transformer architecture, where each layer comprises a multi-head self-attention mechanism and a feed-forward network to capture both local and global dependencies. 

The masked transaction sequence is represented as \( \hat{X}_i^\text{masked} = [\texttt{[CLS]}, tok_1, tok_2, \dots, tok_L] \), where \( L <L_{max} \), \( L \) is the sequence length, \(L_{max}\) is the maximum sequence length allowed by the bert model, \( \texttt{[CLS]} \) is a special token summarizing the sequence. The sequence representations are computed as:
\begin{equation}
\label{eq:seq_encoder}
    \boldsymbol{\tilde{h}}^{(l+1)}_\texttt{CLS}, \boldsymbol{\tilde{h}}^{(l+1)}_1, \ldots, \boldsymbol{\tilde{h}}^{(l+1)}_L = f_\texttt{seq}(\boldsymbol{{h}}^{(l)}_\texttt{CLS}, \boldsymbol{{h}}^{(l)}_1, \ldots, \boldsymbol{{h}}^{(l)}_L),
\end{equation}
where \( l = 1, \ldots, N \) denotes the Transformer layers, and the initial representations \( \boldsymbol{h}^{(0)} \) are derived from token embeddings \( \boldsymbol{e}_j \) and positional embeddings \( \boldsymbol{p}_j \), such that:
\begin{equation}
\label{eq:init_rep}
\boldsymbol{h}^{(0)}_j = \boldsymbol{e}_j + \boldsymbol{p}_j, \quad j \in \{ \texttt{[CLS]}, 1, \ldots, L \}.
\end{equation}
The \( \texttt{[CLS]} \) token serves as a pooling representation for the entire sequence, aggregating the semantic features of transaction records. After passing through \( N \) Transformer layers, the final representation \( \boldsymbol{\tilde{h}}^{(N)}_\texttt{CLS} \) is used as the account's aggregated transaction semantics embedding.

\paragraph{\textbf{Graph Encoder}}

We utilize a graph neural network (GNN) to encode the Transaction Knowledge Graph (TKG). As described in Section~2.3, each entity node is initialized with expert-engineered features that capture statistical and structural characteristics of Ethereum accounts. To further model entity interactions, we apply GNN convolution for information propagation and aggregation \cite{scarselli2008gnn,gat}. The encoding process is defined as:
\begin{equation}
\label{eq:gnn_encoder}
\boldsymbol{\tilde{v}}_1^{(l+1)}, \dots, \boldsymbol{\tilde{v}}_{|V|}^{(l+1)} = f_\texttt{gnn}(\boldsymbol{v}_1^{(l)}, \dots, \boldsymbol{v}_{|V|}^{(l)}),
\end{equation}
where \( l \) represents the layer index, and \( \boldsymbol{v}_i^{(l)} \) is the embedding of node \( i \) at layer \( l \). The update rule follows:
\begin{equation}
\label{eq:node_update}
\boldsymbol{v}_i^{(l+1)} = f_v\left(\sum_{s \in \mathcal{N}_i \cup \{i\}} \alpha_{s,i} \boldsymbol{m}_{si}\right),
\end{equation}
where \( \mathcal{N}_i \) is the neighborhood of node \( i \), \( \alpha_{s,i} \) is the attention weight, and \( f_v \) is a feed-forward network. The message \( \boldsymbol{m}_{si} \) is computed as:
\begin{equation}
\label{eq:message}
\boldsymbol{m}_{si} = f_m(\boldsymbol{v}_s^{(l)}, \boldsymbol{r}_{si}),
\end{equation}
where \( \boldsymbol{r}_{si} \) represents the relation embedding, and \( f_m \) is a linear transformation. The attention weight \( \alpha_{s,i} \) is given by:
\begin{equation}
\label{eq:attention}
\alpha_{s,i} = \text{Softmax}\left(\frac{\boldsymbol{q}_s \boldsymbol{k}_i^\top}{\sqrt{d}}\right),
\end{equation}
where \( \boldsymbol{q}_s = f_q(\boldsymbol{v}_s^{(l)}) \) and \( \boldsymbol{k}_i = f_k(\boldsymbol{v}_i^{(l)}, \boldsymbol{r}_{si}) \) are the query and key vectors obtained through linear transformations.

By aggregating information from relevant neighbors, the GNN encoding process refines entity representations while aligning their dimensions with semantic embeddings. This enables the model to capture both local and global transaction patterns within the TKG.

\subsection{Pre-training KGBERT4Eth}

The encoded textual transaction records and TKG are synergized using the proposed mask-invariant attention synergy method. KGBERT4Eth is pre-trained through the integration of two tasks: Biased Mask Prediction (BMP) and Transaction Link Prediction (TLP).

\paragraph{\textbf{Mask-invariant Attention Synergy}}

To leverage the strengths of language models for sequential data and graph representation learning for relational interactions, we propose an Mask-invariant Attention Synergy (MiAS) method that alleviates knowledge noise and semantic disruption in the pre-train stage. Specifically, we introduce a query-aligned dictionary mechanism to ensure that the semantic embeddings of an account interact only with their corresponding TKG entities and related triplets, avoiding interference from unrelated entities and interactions. Furthermore, To preserve semantic meaning and the integrity of masked tokens during interaction, we adopt an attention visibility matrix inspired by \cite{liu2020k-BERT}. This matrix, implemented as an attention mask, restricts semantic embeddings to interact with entities exclusively through the \texttt{[CLS]} token. This process is formally denoted as:

Let the account transaction embeddings be \( \boldsymbol{\tilde{h}}^{(l)} = [\boldsymbol{\tilde{h}}_\texttt{CLS}^{(l)}, \boldsymbol{\tilde{h}}_1^{(l)}, \dots, \boldsymbol{\tilde{h}}_L^{(l)}] \) and the TKG entity embeddings be \( \boldsymbol{\tilde{v}}^{(l)} = [\boldsymbol{\tilde{v}}_1^{(l)}, \boldsymbol{\tilde{v}}_2^{(l)}, \dots, \boldsymbol{\tilde{v}}_{|V|}^{(l)}] \). Using a multi-head cross-attention mechanism, we compute updated semantic embeddings \( \boldsymbol{\hat{h}}^{(l)} \) and updated entity embeddings \( \boldsymbol{\hat{v}}^{(l)} \). Concretely, the semantic embeddings are updated as:
\begin{equation}
\label{eq:semantic_attention}
\begin{aligned}
    \alpha_{h,v} &= \text{Softmax}\left(\frac{\boldsymbol{q}_h \boldsymbol{k}_v^\top+M}{\sqrt{d}}\right), \\
    \boldsymbol{\hat{h}}^{(l)} &= \sum_{v \in \mathscr{D}(h)} \alpha_{h,v} \cdot \boldsymbol{v}_v,
\end{aligned}
\end{equation}
where \( \boldsymbol{q}_h = f_q(\boldsymbol{\tilde{h}}^{(l)}) \) and \( \boldsymbol{k}_v = f_k(\boldsymbol{\tilde{v}}^{(l)}) \) are the query and key vectors computed via linear transformations \( f_q \) and \( f_k \), respectively. The dictionary \( \mathscr{D} \) ensures that each semantic embedding interacts only with its corresponding entities and related triplets. \( M \) is a visibility matrix that sets the attention scores of all elements in \( \boldsymbol{\tilde{h}}^{(l)} \) except \( \boldsymbol{\tilde{h}}_\texttt{CLS}^{(l)} \) to 0, ensuring that \( \boldsymbol{\tilde{h}}^{(l)} \) enriches account representations without altering the original sentence semantics or disrupting masked tokens.

Similarly, the entity embeddings are updated by treating \( \boldsymbol{\tilde{v}}^{(l)} \) as queries and \( \boldsymbol{\tilde{h}}^{(l)} \) as keys and values, where \( \boldsymbol{q}_v = f_q(\boldsymbol{\tilde{v}}^{(l)}) \) and \( \boldsymbol{k}_h = f_k(\boldsymbol{\tilde{h}}^{(l)}) \) follow similar definitions. This symmetric interaction ensures that both modalities are enriched by the features of the other.

The updated representations, \( \boldsymbol{\hat{h}}^{(l)} \) and \( \boldsymbol{\hat{v}}^{(l)} \), serve as the refined semantic and entity embeddings, which are directly utilized in their respective pre-training tasks. By restricting interactions to relevant entities and enforcing attention visibility constraints, the MiAS aligns semantic and structural embeddings, enabling mutual enhancement of the subsequent pre-training tasks for the language model and graph model.

\paragraph{\textbf{Pre-training Tasks}}
The objective of pre-training is to encourage the model to learn more discriminative representations of account transaction texts while thoroughly grounding and contextualizing the semantic information and interactions within the TKG. To achieve this, we design two self-supervised pre-training tasks: Biased Mask Prediction (BMP) and Transaction Link Prediction (TLP). The final pre-training loss combines the two task-specific losses.

BMP predicts the masked tokens in the transaction text based on their contextual representations. Using the semantic embeddings \( \boldsymbol{h}_i \) obtained from the sequence encoder, the BMP loss for the masked transaction sequence \( \hat{X}_i \) can be formulated as:
\begin{equation}
\label{eq:bmp_loss}
    \mathcal{L}_\text{BMP}(\theta^m) = -\sum_{i=1}^{L} \textup{Mask}(tok_i) \log P(tok_i | \boldsymbol{h}_i),
\end{equation}
where \(tok_i \in \hat{X}_i \), \( P(tok_i | \boldsymbol{h}_i) \) is the softmax probability of the token \( tok_i \) over the entire vocabulary. By leveraging the synergized representations, BMP enriches semantic embeddings with relevant expert structural information, enabling the integration of expert knowledge into semantic representations.

Link prediction is a widely used task in KG representation learning that aims to distinguish valid (positive) triplets from corrupted (negative) triplets. For a triplet \((h, r, t)\) in the TKG, where \( h \), \( r \), and \( t \) represent the head entity, relation, and tail entity respectively, the triplet embeddings are obtained from the graph encoder as \((\boldsymbol{v}_h, \boldsymbol{r}, \boldsymbol{v}_t)\). The TLP loss is defined as:
\begin{equation}
\label{eq:tlp_loss}
\begin{aligned}
    \mathcal{L}_\text{TLP}(\theta^l) = \sum_{(h, r, t) \in S} \Big( - \sigma(d(\boldsymbol{v}_h, \boldsymbol{r}, \boldsymbol{v}_t)) + \\
    \sum_{(h', r, t') \in S'} \sigma(d(\boldsymbol{v}_{h'}, \boldsymbol{r}, \boldsymbol{v}_{t'})) \Big),
\end{aligned}
\end{equation}
where \((h, r, t) \in S \) is the set of positive triplets in the TKG, \((h', r, t') \in S' \) is the set of corrupted triplets generated by replacing either the head or tail entity with a random entity, \( \sigma \) is the sigmoid function (e.g. TransE \cite{bordes2013transe}), and \( d \) is a scoring function that measures the plausibility of a triplet. This task ensures that the graph embeddings effectively capture relational structures within the TKG.

The final pre-training loss, \(\mathcal{L}(\boldsymbol{\theta})\), is optimized over the parameters \(\boldsymbol{\theta} = \{\theta^m, \theta^l\}\) by combining BMP and TLP tasks, expressed as: 
\begin{equation}
\label{eq:combined_loss}
\mathcal{L}(\boldsymbol{\theta}) = \mathcal{L}_\text{BMP}(\theta^m) + \mathcal{L}_\text{TLP}(\theta^l).
\end{equation}

\section{Experimental}
\subsection{Pre-training Data and Implementation Details}
We extracted 3,251,082 Ethereum transaction records involving 496,740 accounts from Etherscan, Kaggle, and Xblock\cite{xblock} to pre-train our model on generic representations of Ethereum accounts and transaction semantics, as well as domain-specific structural knowledge. 
After textualizing the records, we tokenized them using BERT-base’s tokenizer and calculated BM25 scores with the BM25L algorithm. From these records, we built a transaction knowledge graph (TKG) comprising $\sim$500k entities and $\sim$1M edges. During pre-training, the language model's maximum input sequence length $L_{\max}$ was set to 512, and the mask threshold $\tau$ to 0.1. Finetuning datasets employed 2-hop node retrieval with a maximum of 100 nodes. In particular, our model directly processes textual transaction records without building or updating the KG during the fine-tuning and evaluation phases. The downstream data was split into training (70\%), validation (10\%), and test (20\%) sets, and ensure that all accounts that participated in pre-training were excluded from downstream task evaluation, with all results reported on the test set, averaged over five runs. 

\begin{table*}[ht]
    \centering
     \renewcommand{\arraystretch}{1.1}
    \caption{
    Comparison of downstream task performance on the phishing account detection task across three datasets (MultiGraph, B4E, SPN). Metrics include F1(\%), AUC(\%), and FNR(\%) with their respective errors.
    }
    \resizebox{0.97\linewidth}{!}{
    \begin{tabular}{lc|ccccccccc}
    \toprule[1.1pt]
        &   & \multicolumn{3}{c}{MultiGraph} & \multicolumn{3}{c}{B4E} & \multicolumn{3}{c}{SPN} \\ \cmidrule{3-11}
        \multicolumn{2}{c|}{Methods}     & F1 $\uparrow$ & AUC $\uparrow$ & FNR $\downarrow$ & F1 $\uparrow$ & AUC $\uparrow$ & FNR $\downarrow$ & F1 $\uparrow$ & AUC $\uparrow$ & FNR $\downarrow$\\
    \cmidrule(lr){1-2} \cmidrule(lr){3-5} \cmidrule(lr){6-8} \cmidrule(lr){9-11}
    DeepWalk &  & $60.11_{1.12}$ & $77.88_{0.33}$ & $41.44_{1.41}$ & $62.91_{0.88}$ & $78.05_{0.99}$ & $35.44_{0.58}$ & $53.18_{0.95}$ & $74.91_{1.12}$ & $48.72_{0.22}$ \\
    Role2Vec &  & $56.71_{0.96}$ & $77.02_{1.15}$ & $30.66_{0.65}$ & $68.53_{0.55}$ & $79.78_{0.91}$ & $20.11_{1.36}$ & $57.01_{0.49}$ & $74.78_{1.07}$ & $29.80_{1.06}$ \\
    Trans2Vec &  & $68.50_{1.06}$ & $72.56_{1.07}$ & $30.36_{0.42}$ & $36.62_{1.25}$ & $73.44_{0.45}$ & $29.40_{0.86}$ & $52.11_{0.29}$ & $77.57_{0.56}$ & $25.92_{1.36}$ \\
    \midrule
    GCN &  & $43.44_{0.22}$ & $71.87_{0.89}$ & $25.31_{0.33}$ & $62.96_{0.91}$ & $79.54_{0.89}$ & $25.22_{1.02}$ & $48.97_{0.77}$ & $73.79_{1.04}$ & $50.50_{0.82}$ \\
    GAT &  & $41.12_{1.40}$ & $72.92_{1.48}$ & $20.96_{0.49}$ & $61.32_{1.21}$ & $76.84_{0.22}$ & $16.56_{1.31}$ & $60.07_{0.55}$ & $76.68_{0.25}$ & $22.38_{0.69}$ \\
    GSAGE &  & $33.48_{0.95}$ & $73.54_{0.87}$ & $66.74_{0.62}$ & $52.03_{1.13}$ & $74.94_{1.10}$ & $41.98_{1.28}$ & $52.82_{1.01}$ & $77.65_{1.24}$ & $41.41_{0.87}$ \\
    DiffPool &  & $60.30_{0.65}$ & $78.81_{1.42}$ & $42.64_{0.58}$ & $53.08_{0.44}$ & $75.85_{1.03}$ & $49.31_{0.55}$ & $51.63_{1.37}$ & $73.71_{0.64}$ & $48.77_{1.18}$ \\
    U2GNN &  & $59.65_{1.15}$ & $78.55_{0.83}$ & $38.99_{0.85}$ & $60.97_{0.58}$ & $79.91_{1.21}$ & $42.91_{0.49}$ & $54.60_{0.68}$ & $79.96_{0.45}$ & $46.45_{0.67}$ \\
    Graph2Vec &  & $55.14_{1.11}$ & $76.66_{1.44}$ & $56.58_{1.03}$ & $58.26_{0.55}$ & $77.84_{0.73}$ & $52.36_{0.45}$ & $57.76_{1.32}$ & $77.76_{1.12}$ & $55.27_{0.77}$ \\
    TSGN &  & $67.39_{1.05}$ & $72.61_{0.49}$ & $42.60_{0.84}$ & $69.35_{0.96}$ & $73.51_{1.09}$ & $18.79_{1.24}$ & $62.07_{0.92}$ & $78.94_{0.68}$ & $49.17_{0.88}$ \\
    GrabPhisher &  & $78.89_{0.54}$ & $82.02_{0.65}$ & $15.44_{1.03}$ & $67.38_{0.78}$ & $80.19_{0.98}$ & $40.76_{0.76}$ & $73.76_{0.99}$ & $77.69_{1.04}$ & $19.12_{1.10}$ \\
    GAE &  & $44.93_{0.67}$ & $73.67_{1.08}$ & $45.63_{1.12}$ & $50.06_{0.88}$ & $73.81_{1.20}$ & $43.72_{0.91}$ & $36.85_{0.69}$ & $71.56_{0.78}$ & $62.58_{1.22}$ \\
    GATE &  & $45.39_{0.79}$ & $74.90_{1.32}$ & $28.56_{1.11}$ & $58.58_{0.86}$ & $76.11_{1.03}$ & $27.90_{0.59}$ & $68.69_{1.02}$ & $75.61_{0.79}$ & $26.60_{0.45}$ \\
    \midrule
    BERT4ETH &  & $64.20_{0.78}$ & $79.77_{1.11}$ & $26.84_{1.28}$ & $69.05_{1.13}$ & $80.06_{0.92}$ & $38.38_{0.77}$ & $72.98_{0.69}$ & $83.59_{1.04}$ & $32.72_{1.22}$ \\
    ZipZap &  & $65.25_{1.03}$ & $80.87_{1.12}$ & $26.60_{0.45}$ & $68.48_{1.01}$ & $78.68_{1.11}$ & $38.91_{0.92}$ & $72.31_{1.29}$ & $84.82_{0.65}$ & $32.81_{0.99}$ \\
    \midrule
    \cellcolor{Gray}Ours & \cellcolor{Gray} & \cellcolor{Gray}$\textbf{87.60}_{0.65}$ & \cellcolor{Gray}$\textbf{96.01}_{0.49}$ & \cellcolor{Gray}$\textbf{14.11}_{0.87}$ & \cellcolor{Gray}$\textbf{85.43}_{1.22}$ & \cellcolor{Gray}$\textbf{95.34}_{0.92}$ & \cellcolor{Gray}$\textbf{16.03}_{1.13}$ & \cellcolor{Gray}$\textbf{88.02}_{1.02}$ & \cellcolor{Gray}$\textbf{97.12}_{1.15}$ & \cellcolor{Gray}$\textbf{14.75}_{1.09}$ \\
    \bottomrule[1.1pt]
    \end{tabular}}
\label{tab:phish_baseline_with_errors}
\end{table*}

\subsection{Downstream Tasks}
We evaluated the pre-trained model on two Ethereum fraud detection tasks: phishing account detection and account de-anonymization. Phishing attacks are the most common type of fraud on Ethereum. We tested the model's ability to identify phishing accounts using three public datasets: MultiGraph \cite{xblock_1}, B4E \cite{bert4eth}, and SPN \cite{sun2025ethereum}. 
The identities behind blockchain addresses often rely on community labeling or voluntary disclosure by their owners. This anonymity provides protection for various types of financial crimes. To evaluate the model's de-anonymization capability, we obtained transaction data for four types of accounts using labels from GitHub \cite{Brian2023etherscanlabels}: 972 Airdrop Hunters, 178 ICO Wallets, 331 Mev Bots, and 524 Synthetix accounts for evaluation.

\subsection{Baseline Methods}

\paragraph{\textbf{Random Walk-Based}} 
DeepWalk \cite{deepwalk} generates node features through unsupervised random walks without considering node or edge features.  
Role2Vec \cite{role2vec} introduces attribute-based random walks, mapping vertex attribute vectors to roles and learning context probabilities.  
Trans2Vec \cite{trans2vec} is specifically designed for Ethereum fraud detection, employing biased walks based on transaction amounts and timestamps.

\paragraph{\textbf{GNN-Based.}}  
GCN, GAT, and GSAGE \cite{gcn,gat,gsage} are classic neighborhood aggregation algorithms widely used in Ethereum fraud detection.  
DiffPool \cite{diffpool} introduces a differentiable graph pooling module to generate hierarchical graph representations, commonly applied to graph classification tasks.  
U2GNN \cite{U2GNN} uses a self-attention mechanism to capture dependencies between nodes and infer node embeddings through recursive transformations.  
Graph2Vec \cite{graph2vec} and TSGN \cite{tsgn} convert Ethereum node classification tasks into line graph classification problems by sampling subgraphs from target nodes for classification.  
GrabPhisher \cite{grabphisher} dynamically models the temporal features of account transactions and changes in graph topology to identify fraudulent accounts.  
GAE \cite{VGAE} and GATE \cite{gate} learn graph information through graph reconstruction, with GATE incorporating attention mechanisms to focus more on node attribute reconstruction.

\begin{table*}[ht]
    \centering
    \renewcommand{\arraystretch}{1.18}
    \caption{
    Comparison of downstream task performance on the account deanonymization task across different models. Identities include Airdrop Hunter, ICO Wallets, Mev Bot, and Synthetix. Metrics include F1(\%) and FNR(\%) with respective errors.
    }
    \resizebox{1\linewidth}{!}{
    \begin{tabular}{l|cccccccccc}
    \toprule[1.1pt]
        & \multicolumn{2}{c}{Overall} & \multicolumn{2}{c}{Airdrop Hunter} & \multicolumn{2}{c}{ICO Wallets} & \multicolumn{2}{c}{Mev Bot} & \multicolumn{2}{c}{Synthetix} \\ \cmidrule{2-11}
        Methods     & F1 $\uparrow$ & FNR $\downarrow$ & F1 $\uparrow$ & FNR $\downarrow$ & F1 $\uparrow$ & FNR $\downarrow$ & F1 $\uparrow$ & FNR $\downarrow$ & F1 $\uparrow$ & FNR $\downarrow$ \\
    \cmidrule(lr){1-1} \cmidrule(lr){2-3} \cmidrule(lr){4-5} \cmidrule(lr){6-7} \cmidrule(lr){8-9} \cmidrule(lr){10-11}
    DeepWalk & $66.62_{0.48}$ & $34.14_{0.59}$ & $78.86_{1.00}$ & $19.24_{0.57}$ & $14.20_{1.07}$ & $83.55_{1.04}$ & $51.76_{1.03}$ & $49.28_{0.70}$ & $68.81_{0.21}$ & $36.92_{0.43}$ \\
    Role2Vec & $57.36_{1.02}$ & $44.28_{1.39}$ & $72.71_{0.42}$ & $23.25_{1.06}$ & $2.65_{0.56}$ & $96.71_{0.53}$ & $32.37_{1.01}$ & $67.39_{1.06}$ & $57.88_{0.41}$ & $52.74_{0.58}$ \\
    Trans2Vec & $74.91_{0.69}$ & $25.13_{0.61}$ & $85.49_{0.71}$ & $12.94_{1.12}$ & $23.57_{0.63}$ & $76.97_{1.35}$ & $62.50_{0.33}$ & $34.78_{0.74}$ & $78.75_{0.25}$ & $25.32_{0.23}$ \\
    \midrule
    GCN & $59.81_{1.03}$ & $41.18_{0.78}$ & $74.60_{0.75}$ & $22.11_{0.51}$ & $5.68_{1.13}$ & $93.42_{0.23}$ & $35.97_{0.68}$ & $63.77_{0.88}$ & $62.03_{0.37}$ & $46.41_{0.52}$ \\
    GAT & $64.09_{0.54}$ & $36.68_{1.31}$ & $77.68_{1.02}$ & $18.67_{0.37}$ & $8.90_{0.48}$ & $90.13_{0.20}$ & $46.35_{0.27}$ & $52.90_{0.92}$ & $65.29_{0.94}$ & $43.25_{1.27}$ \\
    GSAGE & $55.72_{0.95}$ & $45.80_{1.18}$ & $72.01_{0.72}$ & $24.40_{0.42}$ & $4.10_{0.68}$ & $94.74_{0.97}$ & $29.30_{0.80}$ & $71.01_{1.16}$ & $54.80_{1.20}$ & $54.85_{0.70}$ \\
    DiffPool & $72.34_{0.60}$ & $27.66_{1.09}$ & $84.13_{0.81}$ & $14.09_{0.47}$ & $19.23_{1.08}$ & $80.26_{0.59}$ & $59.03_{0.85}$ & $38.41_{1.25}$ & $76.00_{0.65}$ & $29.54_{0.92}$ \\
    U2GNN & $62.73_{0.70}$ & $38.08_{1.02}$ & $77.80_{0.78}$ & $18.10_{1.23}$ & $5.85_{0.36}$ & $93.42_{0.40}$ & $44.56_{0.67}$ & $54.71_{1.09}$ & $61.56_{0.86}$ & $47.47_{0.79}$ \\
    Graph2Vec & $66.70_{0.89}$ & $33.30_{0.45}$ & $80.31_{1.04}$ & $17.53_{0.62}$ & $12.42_{1.29}$ & $86.84_{0.92}$ & $49.47_{1.22}$ & $49.28_{1.06}$ & $69.97_{0.68}$ & $35.86_{0.59}$ \\
    TSGN & $79.38_{0.65}$ & $20.62_{1.30}$ & $89.74_{1.22}$ & $8.36_{0.42}$ & $36.10_{1.08}$ & $67.11_{0.70}$ & $67.74_{0.48}$ & $31.16_{0.88}$ & $79.44_{1.11}$ & $22.15_{0.92}$ \\
    GrabPhisher & $81.07_{0.78}$ & $18.93_{0.37}$ & $91.93_{0.60}$ & $6.64_{0.72}$ & $39.71_{1.03}$ & $63.82_{0.62}$ & $65.97_{1.16}$ & $31.16_{1.01}$ & $82.03_{1.29}$ & $20.04_{0.54}$ \\
    GAE & $64.26_{1.24}$ & $36.96_{0.87}$ & $77.65_{0.73}$ & $19.82_{1.04}$ & $8.40_{0.74}$ & $90.13_{1.11}$ & $45.94_{0.80}$ & $52.90_{1.28}$ & $66.50_{0.92}$ & $42.19_{0.57}$ \\
    GATE & $66.73_{1.08}$ & $33.86_{0.40}$ & $78.98_{0.99}$ & $18.67_{0.78}$ & $11.70_{1.13}$ & $86.84_{1.08}$ & $52.24_{0.69}$ & $49.28_{1.02}$ & $69.57_{0.88}$ & $35.86_{1.20}$ \\
    \midrule
    BERT4ETH & $76.13_{0.82}$ & $23.44_{1.02}$ & $87.49_{0.54}$ & $10.65_{0.67}$ & $27.87_{1.07}$ & $73.68_{1.10}$ & $63.60_{1.02}$ & $34.78_{1.22}$ & $78.56_{0.70}$ & $24.26_{0.91}$ \\
    ZipZap & $75.91_{0.97}$ & $23.72_{0.66}$ & $85.41_{0.94}$ & $11.80_{0.60}$ & $28.88_{1.14}$ & $73.51_{0.54}$ & $64.75_{1.04}$ & $34.95_{0.98}$ & $79.65_{0.59}$ & $23.21_{1.30}$ \\
    \midrule
    \cellcolor{Gray}Ours & \cellcolor{Gray}$\textbf{90.52}_{0.69}$ & \cellcolor{Gray}$\textbf{8.90}_{0.62}$ & \cellcolor{Gray}$\textbf{98.36}_{0.47}$ & \cellcolor{Gray}$\textbf{0.11}_{0.59}$ & \cellcolor{Gray}$\textbf{65.66}_{0.44}$ & \cellcolor{Gray}$\textbf{42.76}_{0.33}$ & \cellcolor{Gray}$\textbf{84.07}_{0.77}$ & \cellcolor{Gray}$\textbf{17.75}_{0.81}$ & \cellcolor{Gray}$\textbf{88.68}_{0.96}$ & \cellcolor{Gray}$\textbf{9.07}_{0.34}$ \\
    \bottomrule[1.1pt]
    \end{tabular}}
\label{tab:deanrole_baseline_with_errors}
\end{table*}

\paragraph{\textbf{Pre-trained Transformer-Based.}}  
BERT4ETH \cite{bert4eth} is a Transformer specifically designed for Ethereum account representation learning, capturing the inherent dynamic sequential patterns in transactions.  
ZipZap \cite{zipzap} is an improved version of BERT4ETH, employing frequency-aware compression techniques to significantly reduce model size and enhance computational efficiency.

\subsection{Model Performance}
Table~\ref{tab:phish_baseline_with_errors} summarizes the performance of different models on the phishing account detection task across three datasets. All metrics are reported in percentage form without the '\%'. Our proposed model KGBERT4Eth, consistently outperforms the baselines, demonstrating strong capability in handling complex transaction scenarios.

Compared to Random Walk-based and GNN-based methods, KGBERT4Eth shows notable improvements in both F1 and AUC. On the MultiGraph dataset, it attains an F1 of 87.60, exceeding the best Random Walk method (Trans2Vec, 68.50) by 19.10 percentage points and the top GNN method (GrabPhisher, 78.89) by 8.71 points. Moreover, its FNR is just 14.11, underscoring the model’s effectiveness in identifying rare phishing accounts. On the B4E and SPN datasets, KGBERT4Eth maintains similarly strong results, with an F1 of 85.43 and an AUC of 95.34 on B4E, and an F1 of 88.02 and an FNR of 14.75 on SPN.

\begin{table}[t]
    \centering
     \renewcommand{\arraystretch}{1.05}
    \caption{
    Ablation study results on phishing detection (MultiGraph) and deanonymization tasks (Overall). \textit{w/o BMP}: Replaces the BMP strategy with standard random masking.
\textit{w/o TKG}: Completely removes the KG and its expert-curated features, leaving only the LM.
\textit{w/o Expert}: Retains the structural features of the KG but removes expert knowledge and replaced with randomly initialized entity features.
\textit{w/o MiAS}: Uses a simple linear combination instead of the proposed MiAS method.
BERT and RoBERTa: Remove all proposed components and only use BERT and RoBERTa for classification.
    }
    \resizebox{1\linewidth}{!}{
    \begin{tabular}{l|cccc}
    \toprule[1.1pt]
        & \multicolumn{2}{c}{Phisher Detection} & \multicolumn{2}{c}{Deanonymize Identity} \\ \cmidrule{2-5}
        Methods     & F1 $\uparrow$ & FNR $\downarrow$ & F1 $\uparrow$ & FNR $\downarrow$\\
    \cmidrule(lr){1-1} \cmidrule(lr){2-3} \cmidrule(lr){4-5}
    \textit{w/o BMP} & $85.27_{0.83}$ & $17.17_{0.73}$ & $86.67_{0.68}$ & $14.26_{0.97}$ \\
    \textit{w/o TKG} & $86.95_{0.83}$ & $14.84_{1.02}$ & $88.91_{0.50}$ & $12.38_{1.06}$ \\
    \textit{w/o Expert} & $84.49_{0.96}$ & $16.55_{0.48}$ & $88.22_{0.70}$ & $13.62_{0.26}$ \\
    \textit{w/o MiAS} & $86.54_{1.00}$ & $13.72_{0.59}$ & $86.81_{0.85}$ & $14.85_{0.98}$ \\
    BERT & $82.80_{1.11}$ & $17.54_{0.26}$ & $82.73_{0.84}$ & $16.96_{0.32}$ \\
    RoBERTa & $80.92_{1.22}$ & $20.17_{1.37}$ & $83.64_{0.45}$ & $16.11_{0.90}$ \\
    \midrule
    \rowcolor{Gray} Ours & \boldmath{$\textbf{88.02}_{0.58}$} & \boldmath{$\textbf{14.75}_{0.47}$} & \boldmath{$\textbf{90.52}_{0.67}$} & \boldmath{$\textbf{8.90}_{0.59}$} \\
    \bottomrule[1.1pt]
    \end{tabular}}
\label{tab:ablation_study}
\end{table}

\begin{table}[t]
    \centering
    \renewcommand{\arraystretch}{1.05}
    \caption{
    Results of using different score functions in transaction link prediction. Reported on MultiGraph dataset and Overall respectively.
    }
    \resizebox{1\linewidth}{!}{
    \begin{tabular}{l|cccc}
    \toprule[1.1pt]
        & \multicolumn{2}{c}{Phisher Detection} & \multicolumn{2}{c}{Deanonymize Identity} \\ \cmidrule{2-5}
        Score Func     & F1 $\uparrow$ & FNR $\downarrow$ & F1 $\uparrow$ & FNR $\downarrow$\\
    \cmidrule(lr){1-1} \cmidrule(lr){2-3} \cmidrule(lr){4-5}
    DistMult & $87.42_{0.72}$ & $14.22_{0.58}$ & $90.69_{0.65}$ & $8.46_{0.49}$ \\
    QuatE & $87.81_{0.68}$ & $14.64_{0.62}$ & $90.31_{0.69}$ & $8.42_{0.53}$ \\
    TransE & $87.72_{0.71}$ & $14.21_{0.57}$ & $90.22_{0.60}$ & $8.53_{0.48}$ \\
    ComplEx & $87.52_{0.66}$ & $14.50_{0.61}$ & $90.16_{0.62}$ & $8.39_{0.56}$ \\
    \midrule
    \rowcolor{Gray} RotatE (Ours) & \boldmath{$\textbf{88.02}_{0.58}$} & \boldmath{$\textbf{14.75}_{0.47}$} & \boldmath{$\textbf{90.52}_{0.67}$} & \boldmath{$\textbf{8.90}_{0.59}$} \\
    \bottomrule[1.1pt]
    \end{tabular}}
\label{tab:KLP_score_func_comparison}
\end{table}

Transformer-based models generally outperform both Random Walk and GNN-based approaches. In particular, KGBERT4Eth, which interprets account features from a semantic perspective, significantly surpasses BERT4ETH and ZipZap, whose representations rely predominantly on numerical attributes. Across the three phishing datasets, our model achieves an over 15\% improvement in F1 compared to the strongest Transformer-based baselines, illustrating that converting context-independent numerical attributes into meaningful semantic representations enhances the detection of transaction anomalies.

Table~\ref{tab:deanrole_baseline_with_errors} reports the performance of different models on the account deanonymization task, using F1 and FNR for overall performance (Overall) as well as four specific account types: Airdrop Hunter, ICO Wallets, Mev Bot, and Synthetix. 

KGBERT4Eth demonstrates substantial performance advantages in all categories. For Overall, it achieves an F1 of 90.52 and reduces the FNR to 8.90, outperforming all baselines. Specifically, it exceeds the best Random Walk-based method (Trans2Vec, 74.91) by 15.61 percentage points, the top GNN-based method (GrabPhisher, 81.07) by 9.45 points, and the strongest Transformer baseline (BERT4ETH, 76.13) by 14.39 points. Its lower FNR highlights its improved accuracy and robustness across diverse account types. Notably, in the sparse ICO Wallets category, KGBERT4Eth achieves an F1 of 65.66 and an FNR of 42.76, significantly outperforming other methods and demonstrating its ability to handle rare account categories effectively. 

Beyond the overall comparisons, we also examine representative models within each baseline category, on the phishing account detection task. Among the random walk-based approaches, Trans2Vec consistently ranks as the strongest, benefiting from Ethereum-specific sampling strategies. However, its performance exhibits noticeable variance across datasets. While achieving a competitive F1 score on MultiGraph, it struggles to generalize on B4E and SPN, with elevated false negative rates and considerable drops in detection accuracy. In the graph neural network category, GrabPhisher emerges as the most competitive GNN-based baseline, incorporating both temporal and topological information. Nevertheless, despite its relatively strong performance on SPN and MultiGraph, its F1 and AUC scores remain significantly lower than those of KGBERT4Eth, and its detection reliability deteriorates on B4E due to imbalanced class distributions. Other GNN models, such as GCN, GAT, and GSAGE, exhibit high recall but low precision, often producing excessive false positives. These models tend to overfit to structural patterns while failing to incorporate semantic cues necessary for nuanced decision-making. The performance gap is further amplified under class imbalance, where GNNs exhibit limited capacity in distinguishing minority fraudulent classes.

In contrast, Transformer-based models achieve more consistent results, with BERT4ETH and ZipZap outperforming most GNN and graph-based baselines. These models capture sequential patterns in transaction histories, enabling them to better infer semantic irregularities. However, their reliance on numeric feature embeddings without sufficient contextual modeling limits their capacity to generalize across heterogeneous behaviors. 

\subsection{Ablation Study}

\paragraph{\textbf{Effect of Each Proposed Strategy}}

Table~\ref{tab:ablation_study} shows how different pre-training components affect phishing detection and account deanonymization tasks. Removing any single component leads to a performance drop, indicating that each part of our proposed framework is integral to the final outcome. Specifically, omitting the transaction knowledge graph (\textit{w/o TKG}) reduces F1 by about 1.1\% and 1.6\% on the two downstream tasks, respectively. Similarly, excluding the fusion of semantic and TKG embeddings (\textit{w/o MiAS}) lowers F1 by 1.5\% and 6.2\%. Removing the BMP pre-training task (\textit{w/o BMP}) also causes a significant decrease, with F1 scores dropping by around 2.8\% and 3.9\%, highlighting the importance of the BM25-based biased masking prediction task.

We also evaluated two general-purpose language models, BERT and RoBERTa\cite{roberta}, which do not incorporate the biased masking and transaction knowledge graph enhancements. Table~\ref{tab:ablation_study} indicates that both models underperform all ablation variants, showing that combining a Biased Masking Pre-training task with a transaction knowledge graph significantly enhances the language model’s ability to detect transaction anomalies.

\paragraph{\textbf{Different Score Functions in TLP}}

Table \ref{tab:KLP_score_func_comparison} compares the effects of different score functions\cite{zhang2019quaternion,bordes2013transe,trouillon2016complex} in the TLP task. Overall, the RotatE \cite{sun2019rotate} score function demonstrates superior performance in phishing detection, achieving an F1 score of 88.02 and an FNR of 14.75. On the other hand, DistMult \cite{yang2014distMult} performs slightly better in the account deanonymization task, with an F1 score of 90.69 and an FNR of 8.46. However, the performance differences among the models trained with different score functions are not substantial. This suggests that for Ethereum fraud detection, the choice of score functions is less critical when leveraging large-scale pre-trained models.

\paragraph{\textbf{Sensitivity Analysis}} 

\begin{figure}[t]
    \centering     
    \includegraphics[width=1\linewidth]{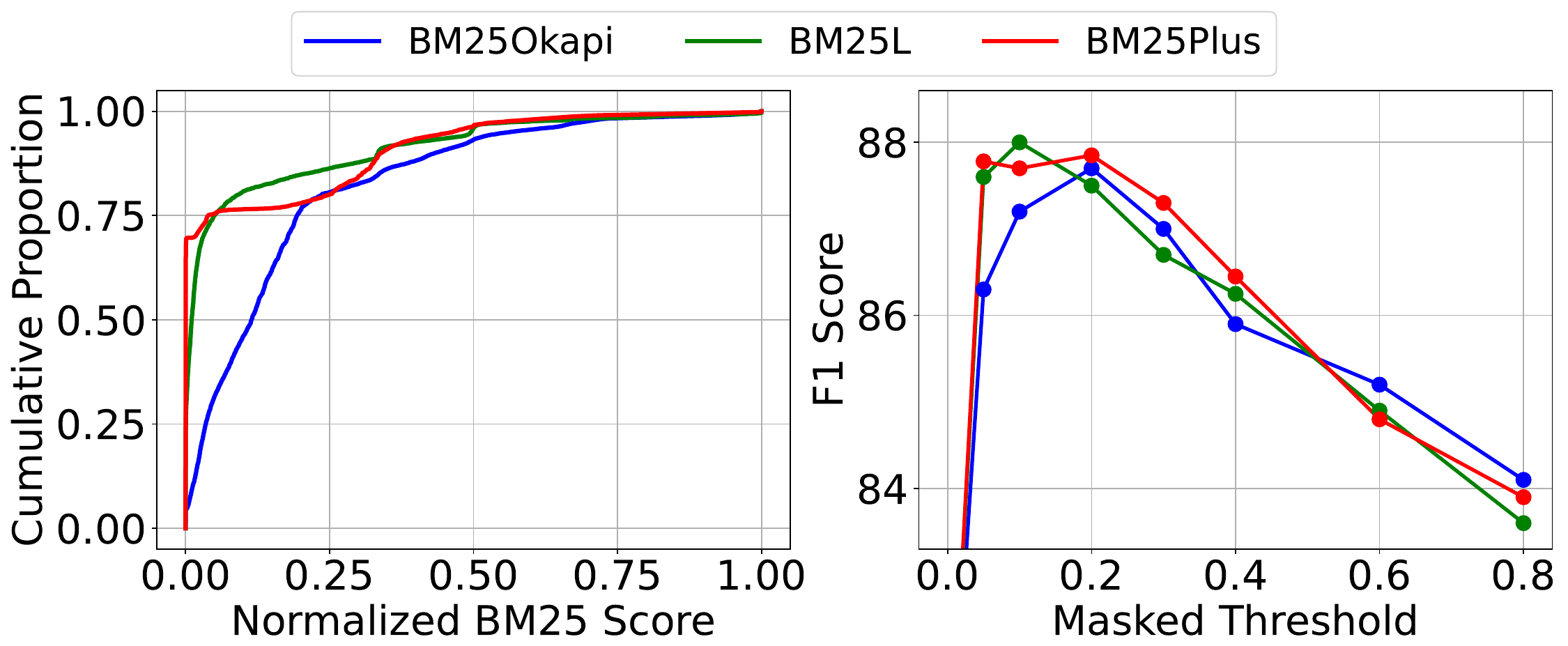}
    
    \caption{Hyperparameter \(\tau \) evaluation using SPN datasets under phishing detection tasks. \emph{left}: The distribution of scores computed by different BM25 algorithms. \emph{right}: how the model performance varies with \( \tau \).}
    
    \label{fig:hyperparam_bm25}
\end{figure}
\begin{figure}[t]
    \centering     
    \includegraphics[width=1\linewidth]{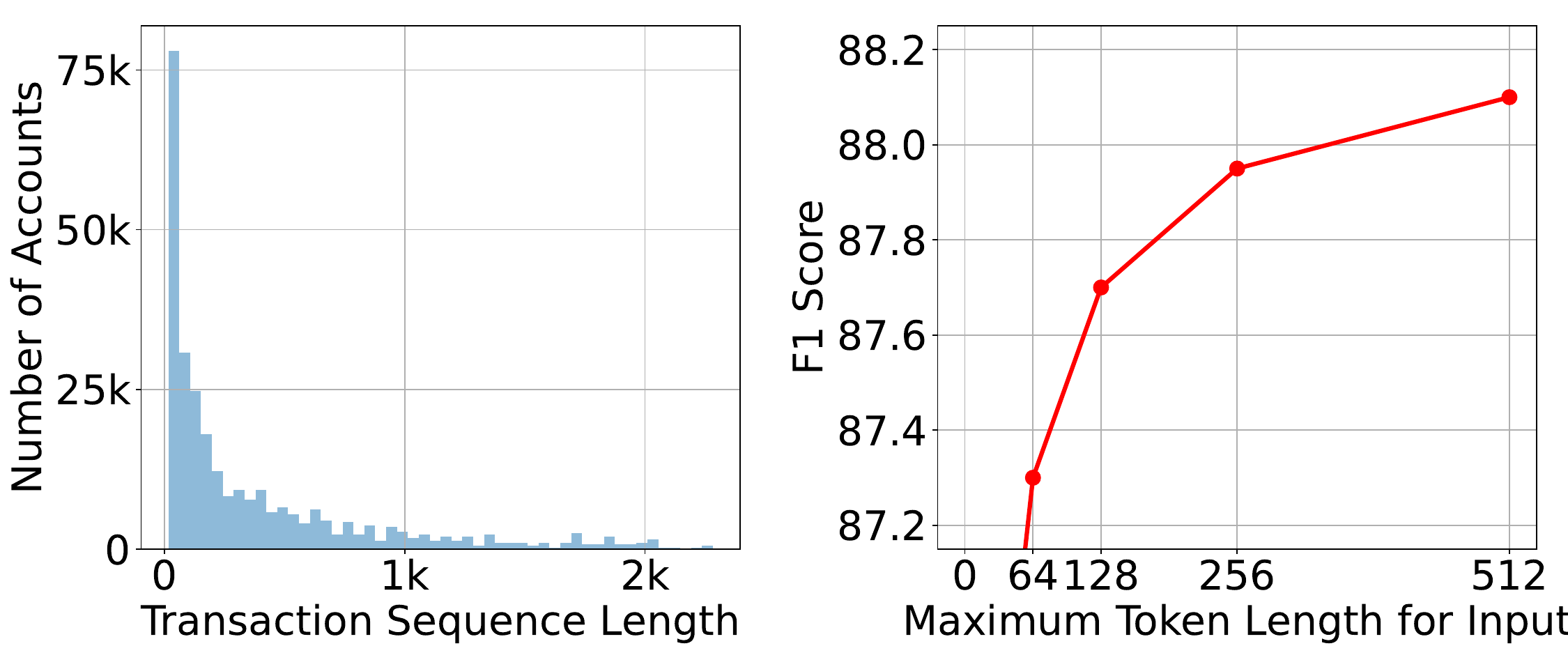}
    \caption{Hyperparameter \( L_{\max} \) evaluation using SPN datasets under phishing detection tasks. \emph{left}: The distribution of token sequence lengths after tokenizing the account transaction texts. \emph{right}: the model performance as a function of \( L_{\max} \) .}
    \label{fig:hyperparam_maxlen}
\end{figure}

We conducted a sensitivity analysis on two hyperparameters, \( \tau \) and \( L_{\max} \), during the pre-training phase to evaluate their individual impacts on model performance. Figure \ref{fig:hyperparam_bm25} (\emph{left}) illustrates the transaction token scores computed using various BM25 algorithms, which exhibit a power-law distribution. After normalization, the majority of scores remain low, with approximately the top 20\% of BM25 scores being filtered out when \( \tau \) is set to 0.2. As shown in Figure \ref{fig:hyperparam_bm25} (\emph{right}), maintaining \( \tau \) within a range of 0.05 to 0.2 yields optimal model performance. However, when \( \tau \) exceeds 0.3, the model's performance deteriorates significantly, irrespective of the BM25 algorithm applied. This decline can be attributed to an insufficient masking ratio of transaction tokens, which hinders the model from effectively learning discriminative features through the masked prediction task.

Figure \ref{fig:hyperparam_maxlen} (\emph{left}) depicts the token sequence length distribution after tokenizing account transaction texts, which also follows a power-law distribution. Approximately 70\% of the account token sequences have lengths shorter than 512, corresponding to around 15 transaction records based on Equation 1. This observation suggests that over 70\% of Ethereum accounts involve fewer than 15 transactions. In Figure \ref{fig:hyperparam_maxlen} (\emph{right}), we observe a consistent improvement in model performance as \( L_{\max} \) increases. Due to the architectural constraints of BERT, we were unable to evaluate the model with input lengths exceeding 512 tokens. Nonetheless, we speculate that allowing larger input lengths could further enhance the model's performance.

\section{Conclusion}
In this work, we proposed KGBERT4Eth, a feature-complete pre-training framework for Ethereum fraud detection that integrates semantic, structural, and expert-defined knowledge in a unified architecture. By coupling a Transaction Language Model (TLM) with a Transaction Knowledge Graph (TKG), our method jointly captures contextual semantics from transaction sequences and structured relationships from the Ethereum transaction network. The introduction of a biased masked prediction objective enables the TLM to focus on statistically anomalous records, while the TKG leverages link prediction to model latent transaction associations and incorporate domain-specific expert features. Furthermore, the mask-invariant attention coordination mechanism ensures stable cross-modal feature fusion without compromising the semantic learning objectives of the language model. Extensive experiments across phishing account detection and de-anonymization benchmarks demonstrate that KGBERT4Eth consistently outperforms state-of-the-art baselines, achieving absolute F1-score improvements of 8–16\% on three phishing detection datasets and 6–26\% on four de-anonymization datasets. In future work, we plan to extend KGBERT4Eth with long-context language models to better capture extended transactional histories, and explore model compression techniques for lightweight deployment in resource-constrained environments.
\bibliographystyle{IEEEtran}
\bibliography{tkde-template}
\end{sloppypar}

\end{document}